\documentclass[twocolumn,showpacs,preprintnumbers,amsmath,amssymb]{revtex4}
\usepackage[dvips]{graphicx}
\usepackage{epsfig}
\usepackage{soul}

\begin{document}

\title{Distillability sudden death in qutrit-qutrit systems under global decoherence}
\author{Mazhar Ali}
\affiliation{Fachbereich Physik, Universit\"{a}t Siegen, 57068, Germany}

\begin{abstract}
Recently Song {\it et\,al}., Phys. Rev. A {\bf 80}, 012331 (2009), have discovered that certain two-qutrit entangled states interacting with multi-local decoherence undergo distillability sudden death whereas their locally equivalent states do not exhibit such behavior. We have found that their basic equation of motion missed some terms and therefore the subsequent calculations and conclusions are not correct/valid except that distillability sudden death may occur. We provide the missing terms and point out the corrections which must be done in the literature to remove confusion. Moreover, we have studied this feature for certain entangled states under global, collective and multi-local decoherence. We have found that entanglement sudden death and distillability sudden death may happen under global and multi-local noises. Moreover, local unitary transformations can not avoid distillability sudden death.
\end{abstract}


\maketitle

\section{Introduction}

Quantum entanglement is a vital resource for the field of quantum information sciences \cite{Nielsen2000}. In a noisy environment, it is important to consider possible degradation due to decoherence of any initially prepared entanglement. Recently, Yu and Eberly \cite{YE-PRB-2003, Yu-Eberly} investigated the dynamics of two-qubit entanglement under various types of decoherence. They found that the decay of a single qubit coherence can be slower than the decay of qubit entanglement. The abrupt disappearance of entanglement at a finite time was named ``entanglement sudden death" (ESD). Clearly, such finite-time disentanglement can seriously affect applications of entangled states in quantum information processing. The initial report of ESD in two-qubit entangled states in a specific model was later explored in wider contexts and in higher dimensions of Hilbert space when the qubits are replaced by qutrits or qudits \cite{ESD-2007, ESD-2008, AJ-PRA76-2007, Ann-Jaeger, AJ-PRB, AJ-JMO-2007, Rau, ESD-2009}. The experimental evidences for this effect have been reported for optical setups \cite{Exp1-ESD} and atomic ensembles \cite{Exp2-ESD}.

For qubit-qubit and qubit-qutrit systems Peres-Horodecki criterion divides the set of quantum states into entangled and separable states \cite{Pe-PRL96, HHH-PLA96}. However for higher dimensional bipartite systems such characterization is not easy \cite{Horodecki-RMP-2009}. Nevertheless, bipartite entangled states can be divided into free-entangled states and bound-entangled states \cite{Horodecki-RMP-2009, Horodecki-PRL80-1998}. Free-entangled states can be distilled under local operations and classical communication (LOCC) whereas bound-entangled states can not be distilled to pure-state entanglement no matter how much copies are available. Bound entanglement itself can not be used for quantum information processing, however it may activate teleportation fidelity \cite{Horodecki-PRL82-1999}. Bound entanglement manifests the irreversibility in asymptotic manipulation of entanglement \cite{Yang-PRL95-2005}. Bound-entanglement constructed by purely mathematical arguments have existence in physical processes \cite{Toth-PRL99-2007}.

Recently, it was proposed that free-entangled states may be converted to bound-entangled states in the presence of local decoherence \cite{Song-PRA80-2009}. In particular, it was shown that certain free entangled states of qutrit-qutrit systems become non-distillable in a finite time under the influence of local decoherence. Such behavior has been named as distillability sudden death (DSD). We have found that the calculations done in \cite{Song-PRA80-2009} are not correct. Nevertheless, the main idea is still working, i.\,e., there are certain free-entangled quantum states which dynamically convert to bound-entangled states. We have rectified the equation of motion for the physical system and pointed out the possible cause of confusion. In addition, we have studied this problem for more general decoherence, i.\,e., when the qutrit-qutrit system is subjected to both collective and multi-local noises. We have found that distillability sudden death may occur under multi-local and global decoherence. However, if qutrit-qutrit states are subjected only to collective noise then there are certain states which never undergo distillability sudden death and entanglement sudden death.

This paper is organized as follows. In Sec. \ref{Sec:Model}, we discuss the physical model and the basic equation of motion along with its solution. We point out the corrections to be made in the published literature. We discuss briefly the idea of distillability sudden death. In Sec. \ref{Sec:Results}, we demonstrate the possibility of distillability sudden death for both global decoherence and multi-local decoherence. Finally, we conclude our work in Sec. \ref{Sec:Conc}.

\section{Dynamics of qutrit-qutrit system under global decoherence} \label{Sec:Model}

\begin{figure}[h]
\scalebox{1.6}{\includegraphics[width=1.4in]{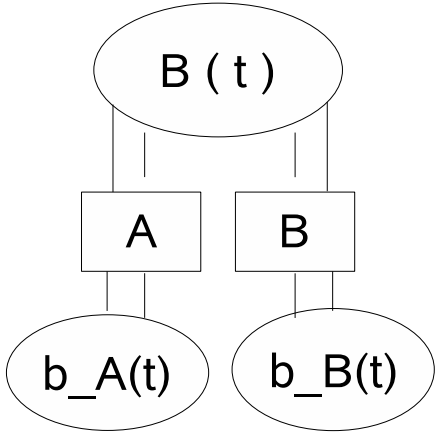}}
\caption{Two qutrits $A$ and $B$ are collectively interacting with the stochastic magnetic field $B(t)$ and separately interacting with the stochastic magnetic fields $b_A(t)$ and $b_B(t)$.}
\label{Fig:model}
\end{figure}

Our physical model consists of two qutrits (two-three level atoms for example) $A$ and $B$ that are coupled to a noisy environment both singly and collectively as shown in Figure \ref{Fig:model}.
The Hamiltonian of the system ($\hbar = 1$) can be written as \cite{AJ-JMO-2007} 
\begin{eqnarray}
H(t) = - \frac{\mu}{2} \, \big[ \, b_A(t) \, \sigma_z^A + b_B(t) \, \sigma_z^B + B(t) (\sigma_z^A + \sigma_z^B) \,  \big] \, , \label{Eq:Ham} 
\end{eqnarray}
where $\mu$ is gyromagnetic ratio and $\sigma_z $ denotes the dephasing operator for qutrits $A$ and $B$. The stochastic magnetic fields refer to statistically independent classical Markov processes satisfying the conditions
\begin{eqnarray} 
\langle b_i(t) \, b_i(t')\rangle &=& \frac{\Gamma_1}{\mu^2} \, \delta(t-t') \,, \, \langle B(t) \, B(t')\rangle = \frac{\Gamma_2}{\mu^2} \, \delta(t-t') \,, \nonumber \\&& \langle b_i(t)\rangle = 0 \, , \quad \langle B(t)\rangle = 0 \, ,
\end{eqnarray}
with $\langle \cdots \rangle$ as ensemble time average, $\Gamma_1$ and $\Gamma_2$ denote the phase-damping rates for multi-local and collective decoherence, respectively.

Let $|2\rangle$, $|1\rangle$, and $|0\rangle$ be the first excited state, second excited, and ground state of the qutrit, respectively. We choose the basis $ \{ \, |2,2\rangle$, $|2,1\rangle$, $|2,0\rangle$, $|1,2\rangle$, $|1,1\rangle$, $|1,0\rangle$, $|0,2\rangle$, $|0,1\rangle$, $|0,0\rangle \, \}$. The time-dependent density matrix for two-qutrit system is obtained by taking ensemble average over the three noise fields, i.\,e., $\rho(t) = \langle\langle\langle\rho_{st}(t)\rangle\rangle\rangle$, where $\rho_{st}(t) = U(t) \rho(0) U^\dagger(t)$ and $U(t) = \exp[-\mathrm{i} \int_0^t \, dt' \, H(t')]$. The dynamics of density matrix can be given by operator sum representation \cite{AJ-PRA76-2007, AJ-JMO-2007} as $\rho(t) = \sum_j^n \, K_j^\dagger(t) \rho(0) K_j(t)$, where $K_j$ are Kraus operators \cite{Kraus} that preserve the positivity and unit trace conditions, i.\,e., $\sum_j^n \, K_j^\dagger K_j = \mathbb{I}$. The most general solution of $\rho(t)$ interacting with global (multi-local + collective) decoherence under the assumption that the system is not initially correlated with any of the three environments is given as
\begin{eqnarray}
\rho(t) = \sum_{i,j=1}^3 \, \sum_{k=1}^3 \, (D_k^{AB \, \dagger} \, F_j^{B \,\dagger} \, E_i^{A \, \dagger}) \, \rho(0) \, (E_i^A \, F_j^B \, D_k^{AB})\, , \label{Eq:GS}
\end{eqnarray}
where the terms describing the interaction with the local magnetic fields $b_A(t)$ and $b_B(t)$ are $E_1^A = \mathrm{diag} (1, \, \gamma_A(t), \, \gamma_A(t)) \otimes \mathbb{I}_3$, $E_2^A = \mathrm{diag} (0,\,\omega_A(t), \,0) \otimes \mathbb{I}_3$, $E_3^A = \mathrm{diag}(0, \, 0, \, \omega_A(t))\otimes \mathbb{I}_3$, $F_1^B = \mathbb{I}_3 \otimes \mathrm{diag} (1, \, \gamma_B(t), \, \gamma_B(t))$, $F_2^B = \mathbb{I}_3 \otimes \mathrm{diag} (0, \, \omega_B(t), \, 0)$, and $F_3^B = \mathbb{I}_3 \otimes \mathrm{diag} (0, \, 0, \, \omega_B(t))$. The terms describing interaction with collective magnetic field $B(t)$ involve the operators $D_1^{AB} = \mathrm{diag} (\gamma(t), 1,1,1,\gamma(t) \,,1,1,1,\gamma(t))$, $D_2^{AB} = \mathrm{diag} (\omega_1(t), 0,0,0,\omega_2(t) \,,0,0,0,\omega_2(t))$, $D_3^{AB} = \mathrm{diag} (0,0,0,0,\omega_3(t),0,0,0,\omega_3(t))$. The time dependent parameters are defined as $\gamma_A = \gamma_B = \mathrm{e}^{-\Gamma_1 t/2}$, $\omega_A(t) = \sqrt{1-\gamma_A^2(t)}$, $\omega_B(t) = \sqrt{1-\gamma_B^2(t)}$, $\gamma = \mathrm{e}^{-\Gamma_2 t/2}$, $\omega_1(t) = \sqrt{1-\gamma^2(t)}$, $\omega_2(t) = - \gamma^2(t) \, \sqrt{1-\gamma^2(t)}$, and $\omega_3(t) = (1-\gamma^2(t)) \, \sqrt{1+\gamma^2(t)}$.

Eq.~(\ref{Eq:GS}) contains $27$ operators of the form, $G_1 = E_1 \, F_1 \, D_1$, $G_2 = E_1 \, F_1 \, D_2$,$\ldots$, $G_{27} = E_3 \, F_3 \, D_3$, and indeed they satisfy the condition $\sum_{n=1}^{27} \, G_n^\dagger \, G_n = \mathbb{I}_9$. The matrix form of Eq.~(\ref{Eq:GS}) is given as
\begin{widetext}
\begin{eqnarray}
\rho(t) = \left(
\begin{array}{lllllllll}
\rho_{11} & \gamma \gamma_B \rho_{12} & \gamma \gamma_B \rho_{13} & \gamma \gamma_A \rho_{14} & \gamma^4 \gamma_A \gamma_B \rho_{15} & \gamma \gamma_A \gamma_B \rho_{16} & \gamma \gamma_A \rho_{17} & \gamma \gamma_A \gamma_B \rho_{18} & \gamma ^4 \gamma_A \gamma_B \rho_{19} \\
\gamma \gamma_B \rho_{21} & \rho_{22} & \gamma_B^2 \rho_{23} & \gamma_A \gamma_B \rho_{24} & \gamma \gamma_A \rho_{25} & \gamma_A \gamma_B^2 \rho_{26} & \gamma_A \gamma_B \rho_{27} & \gamma_A \rho_{28} & \gamma \gamma_A \gamma_B^2 \rho_{29} \\
\gamma \gamma_B \rho_{31} & \gamma_B^2 \rho_{32} & \rho_{33} & \gamma_A \gamma_B \rho_{34} & \gamma \gamma_A \gamma_B^2 \rho_{35} & \gamma_A \rho_{36} & \gamma_A \gamma_B \rho_{37} & \gamma_A \gamma_B^2 \rho_{38} & \gamma \gamma_A \rho_{39} \\
\gamma \gamma_A \rho_{41} & \gamma_A \gamma_B \rho_{42} & \gamma_A \gamma_B \rho_{43} & \rho_{44} & \gamma \gamma_B \rho_{45} & \gamma_B \rho_{46} & \gamma_A^2 \rho_{47} & \gamma_A^2 \gamma_B \rho_{48} & \gamma \gamma_A^2 \gamma_B \rho_{49} \\
\gamma^4 \gamma_A \gamma_B \rho_{51} & \gamma \gamma_A \rho_{52} & \gamma \gamma_A \gamma_B^2 \rho_{53} & \gamma \gamma_B \rho_{54} & \rho_{55} & \gamma \gamma_B^2 \rho_{56} & \gamma \gamma_A^2 \gamma_B \rho_{57} & \gamma \gamma_A^2 \rho_{58} & \gamma_A^2 \gamma_B^2 \rho_{59} \\
\gamma \gamma_A \gamma_B \rho_{61} & \gamma_A \gamma_B^2 \rho_{62} & \gamma_A \rho_{63} & \gamma_B \rho_{64} & \gamma \gamma_B^2 \rho_{65} & \rho_{66} & \gamma_A^2 \gamma_B \rho_{67} & \gamma_A^2 \gamma_B^2 \rho_{68} & \gamma \gamma_A^2 \rho_{69} \\
\gamma \gamma_A \rho_{71} & \gamma_A \gamma_B \rho_{72} & \gamma_A \gamma_B \rho_{73} & \gamma_A^2 \rho_{74} & \gamma  \gamma_A^2 \gamma_B \rho_{75} & \gamma_A^2 \gamma_B \rho_{76} & \rho_{77} & \gamma_B \rho_{78} & \gamma \gamma_B \rho_{79} \\
\gamma \gamma_A \gamma_B \rho_{81} & \gamma_A \rho_{82} & \gamma_A \gamma_B^2 \rho_{83} & \gamma_A^2 \gamma_B \rho_{84} & \gamma \gamma_A^2 \rho_{85} & \gamma_A^2 \gamma_B^2 \rho_{86} & \gamma_B \rho_{87} & \rho_{88} & \gamma \gamma_B^2 \rho_{89} \\
\gamma ^4 \gamma_A \gamma_B \rho_{91} & \gamma \gamma_A \gamma_B^2 \rho_{92} & \gamma \gamma_A \rho_{93} & \gamma \gamma_A^2 \gamma_B \rho_{94} & \gamma_A^2 \gamma_B^2 \rho_{95} & \gamma \gamma_A^2 \rho_{96} & \gamma \gamma_B \rho_{97} & \gamma \gamma_B^2 \rho_{98} & \rho_{99}
\end{array}
\right) \label{Eq:MF}
\end{eqnarray}
\end{widetext}
We note that if multi-local fields are turned off, i.\,e., $\gamma_A = \gamma_B = 1$, and qutrit-qutrit system is subjected to collective noise, then decoherence free subspaces (DFS) \cite{YE-PRB-2003, AJ-JMO-2007} do appear in this system. Similarly, we can concentrate on the dynamics only for multi-local decoherence by turning off the collective magnetic field, i.\,e., $\gamma = 1$ in Eq.~(\ref{Eq:MF}). 

We point out here that Eq.~($1$) of \cite{Song-PRA80-2009} is not correct as the authors studied the case of multi-local noise (with $\gamma = 1$) and defined the general solution as $\rho(t) = \sum_{i,j = 1}^2 \, F_j^\dagger(t) \, E_i^\dagger(t) \rho(0) \, E_i \, F_j$ (compare it with Eq.~(\ref{Eq:GS})). This summation contain four terms instead of nine. They missed the matrices $E_3$ and $F_3$. In fact the equation ($\rho(t) = \sum_{i,j = 1}^2 \, F_j^\dagger(t) \, E_i^\dagger(t) \rho(0) \, E_i \, F_j$) was written in \cite{AJ-PRA76-2007}, where the authors studied the existence of entanglement sudden death for qudit-qudit ($d \otimes d$) systems. In that case, the solution must be $\rho(t) = \sum_{i,j = 1}^d \, F_j^\dagger(t) \, E_i^\dagger(t) \rho(0) \, E_i \, F_j$ and the corresponding operators must be $E_1(t) = \mathrm{diag} (1, \gamma_A, \gamma_A, \ldots,\gamma_A) \otimes \mathbb{I}_d$, $E_2(t) = \mathrm{diag} (0,\omega_A,0,\ldots,0)\otimes \mathbb{I}_d$,$\ldots$,$E_d(t) = \mathrm{diag}(0,0,0,\ldots,\omega_A) \otimes \mathbb{I}_d$, and $F_1(t) = \mathbb{I}_d \otimes \mathrm{diag} (1, \gamma_B, \gamma_B, \ldots,\gamma_B) $, $F_2(t) = \mathbb{I}_d \otimes \mathrm{diag} (0,\omega_B,0,\ldots,0) $,$\ldots$,$F_d(t) = \mathbb{I}_d \otimes \mathrm{diag}(0,0,0,\ldots,\omega_B)$. It is quite straight forward to show that with $E_i$ and $F_j$ defined in \cite{AJ-PRA76-2007}, the partial transpose of isotropic states always contain one negative eigenvalue and the conclusions that isotropic states always exhibit entanglement sudden death would not be correct. However, the conclusions of Ref.\,\cite{AJ-PRA76-2007} are correct and we assume that Ann-Jaeger only wrote the wrong matrices but worked with correct ones. The proof of this assumption can be found in Ref.\,\cite{AJ-JMO-2007}, where again Ann-Jaeger wrote wrong operators in their Section $2$, however they provide the correct matrix as Eq.~($2$) of Ref.\,\cite{AJ-JMO-2007}, which validates their conclusions. However, such confusion should be rectified for broader interest of readers. In fact, the matrices $E_1$, $E_2$, and $E_3$ for a qutrit were used by Ann-Jaeger in their study of entanglement sudden death in qubit-qutrit systems \cite{Ann-Jaeger}. Song {\it et\,al}., \cite{Song-PRA80-2009} probably have taken the results of Ref.\,\cite{AJ-PRA76-2007} and discovered the very interesting feature of distillability sudden death for qutrit-qutrit systems. As a result of this confusion, Eq.~($3$) of Ref.\,\cite{Song-PRA80-2009} is wrong. However, as we demonstrate in Sec. \ref{Sec:Results} that some of their conclusions still survive, in particular, free-entangled states undergo distillability sudden death. However, the local unitary transformation can not avoid distillability sudden death.

\section{Distillability sudden death} \label{Sec:Results}

Before demonstrating the possibility of distillability sudden death, we discuss the characterization of bound-entangled states. It was proved \cite{Horodecki-PRL80-1998} that bound-entangled states have positive partial transpose (PPT) and hence non-distillable under LOCC. There is not a unique criterion to detect bound-entangled states. Even for qutrit-qutrit system, there are various bound-entangled states and a single criterion is not capable to detect all of them (see for example \cite{Clarisse} and references therein). However, one can use the realignment criterion \cite{Chen-QIQ-2003} to detect certain bound-entangled states. The realignment of a given density matrix is obtained as $(\rho^R)_{ij,kl} = \rho_{ik,jl}$. For a separable state $\rho$, realignment criterion implies that $\|\rho^R\| \leq 1$. For a PPT-state, the positive value of the quantity $\|\rho^R\|-1$ can prove the bound-entangled state.

Let us consider a particular initial state given as
\begin{eqnarray}
\rho_\alpha = \frac{2}{7} \, |\Psi_+\rangle\langle\Psi_+| + \frac{\alpha}{7} \, \sigma_+ + \frac{5-\alpha}{7} \, \sigma_-\,, \label{Eq:H-BE}
\end{eqnarray}
where $2 \leq \alpha \leq 5$, the separable states $\sigma_+ = 1/3 (| 0,1 \rangle\langle 0,1 | + |1,2 \rangle\langle 1,2 | + | 2,0 \rangle\langle 2,0 |)$, and $\sigma_- = 1/3 (| 1,0 \rangle\langle 1,0 | + |2,1 \rangle\langle 2,1 | + | 0,2 \rangle\langle 0,2 |)$, and $|\Psi_+\rangle = 1/\sqrt{3} (|0,0\rangle + |1,1\rangle + |2,2\rangle)$ a maximally entangled state. It was shown \cite{Horodecki-PRL82-1999} that $\rho_\alpha$ is separable for $2 \leq \alpha \leq 3$, bound-entangled for $3 < \alpha \leq 4$, and free-entangled for $4 < \alpha \leq 5$. It was claimed \cite{Song-PRA80-2009} that the partial transpose of time-dependent Eq.~(\ref{Eq:H-BE}) in the range $4 < \alpha \leq 5$ always has a negative eigenvalue in finite time. We have already mentioned that this claim is a result of an incorrect equation of motion. The fact is that the possible negative eigenvalues of partial transpose of Eq.~(\ref{Eq:H-BE}) are given as
\begin{eqnarray}
\lambda_1 &=& \frac{1}{42} \left(5 - \sqrt{16 \, \gamma^8 \, \gamma_A^2 \, \gamma_B^2 + 4 \, \alpha^2 - 20 \, \alpha + 25}\right)\,,\nonumber \\
\lambda_2 &=& \frac{1}{42} \left(5 - \sqrt{16 \, \gamma^8 \, \gamma_A^2 \, \gamma_B^2 + 4 \, \alpha^2 - 20 \, \alpha + 25 }\right), \nonumber \\
\lambda_3 &=& \frac{1}{42} \left(5 - \sqrt{16 \gamma_A^4 \, \gamma_B^4 + 4 \, \alpha^2 - 20 \, \alpha + 25} \right). \label{Eq:ev1}
\end{eqnarray}
These eigenvalues obviously become positive at a finite time. It can be observed that if we turn off the multi-local fields and turn on only collective interaction, then these states always has a negative value, which implies no entanglement sudden death and no distillability sudden death. However, in the presence of global decoherence or multi-local decoherence, entanglement sudden death and distillability sudden death may happen as we demonstrate below.

\begin{figure}[h]
\scalebox{1.9}{\includegraphics[width=1.7in]{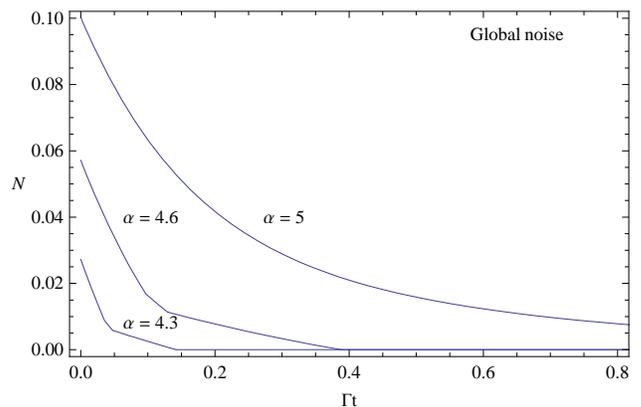}}
\caption{The negativity is plotted against the decay parameter $\Gamma t$ for $\alpha = 4.3$, $\alpha = 4.6$, and $\alpha = 5$.}
\label{Fig:n-hbe}
\end{figure}

The negativity \cite{Vidal-PRA65-2002} for this state can be easily calculated as
\begin{eqnarray}
N(\rho_\alpha(t)) = \max [0, -\lambda_1] + \max[0,-\lambda_2] + \max[0, - \lambda_3]\,. \label{Eq:n-hbe}
\end{eqnarray}
The negativity $N(\rho_\alpha(t))$ is plotted for various values of $\alpha$ in Figure \ref{Fig:n-hbe}, where we have taken $\Gamma_1 = \Gamma_2 = \Gamma$ for ease. Figure \ref{Fig:n-hbe} shows that except for $\alpha = 5$, the negativity becomes zero at finite times leading to a PPT state which might be entangled. The state with $\alpha = 5$ becomes PPT only at infinity, which is definitely a separable state.

\begin{figure}[h]
\scalebox{1.9}{\includegraphics[width=1.8in]{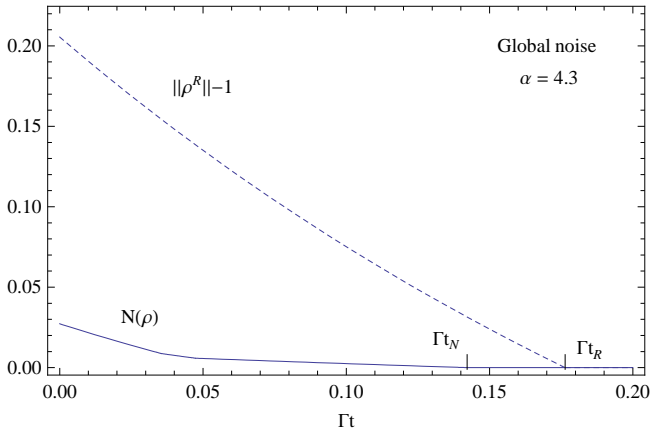}}
\caption{The negativity $N(\rho)$ and the realignment criterion $\|\rho^R\|-1$ is plotted against the decay parameter $\Gamma t$ for $\alpha = 4.3$. The negativity is zero at time $\Gamma t_N \approx 0.1422$, nevertheless, the state is bound-entangled in the range $0.1422 \lesssim \Gamma t \lesssim 0.1764$.}
\label{Fig:DSD-hbe1}
\end{figure}

Let us now demonstrate that the time evolved density matrix $\rho_\alpha(t)$ do undergo distillability sudden death. This could be most easily demonstrated by checking entanglement of $\rho_\alpha(t)$ after it becomes PPT. This relation is plotted in Figure \ref{Fig:DSD-hbe1}, which shows that under global decoherence an initial free-entangled state becomes bound-entangled at a time $\Gamma t_N \approx 0.1422$. The entanglement of PPT-state is verified by the positive value of $\|\rho^R(t)\|-1 $ in the range $0.1422 \lesssim \Gamma t \lesssim 0.1764$. However, the realignment criterion fails to detect the possible entanglement after time $\Gamma t_R \approx 0.1764$. To conclude the separability or entanglement after this time, one could extract three ''$2 \otimes 2$'' density matrices spanned by $\{\, |1,1\rangle, |1,0 \rangle, |0,1 \rangle, |0,0 \rangle \, \}$, $\{\, |2,2\rangle, |2,1 \rangle, |0,2 \rangle, |0,1 \rangle \, \}$, and $\{\, |2,2\rangle, |2,0 \rangle, |1,2 \rangle, |1,0 \rangle \, \}$ from $\rho_\alpha(t)$ and check their properties \cite{Song-PRA80-2009}.

\begin{figure}[h]
\scalebox{1.9}{\includegraphics[width=1.8in]{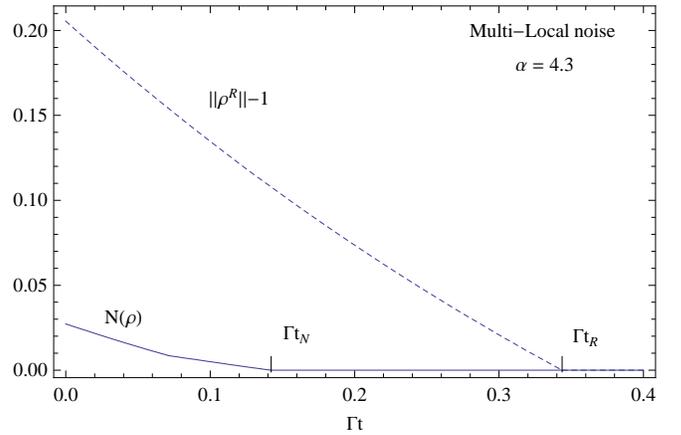}}
\caption{The negativity and the realignment criterion is plotted against the decay parameter $\Gamma t$ for $\alpha = 4.3$. The state becomes bound-entangled at time $\Gamma t_N \approx 0.1422$ and it remains bound-entangled at least in the range $0.1422 \lesssim \Gamma t \lesssim 0.3437$.}
\label{Fig:DSD-hbe2}
\end{figure}

We could turn off the collective field, i.\,e., $\gamma = 1$ and observe the distillability sudden death under multi-local decoherence. Figure \ref{Fig:DSD-hbe2} shows that $\rho_\alpha(t)$ loses its negativity at a finite time $\Gamma t_N \approx 0.1422$ and becomes bound-entangled as the realignment criterion has a positive value for the interval $0.1422 \lesssim \Gamma t \lesssim 0.3437$. Hence $\rho_\alpha(t)$ experiences distillability sudden death under multi-local decoherence as well. It is surprising to note that the dynamics of negativity is not much affected by turning off the collective field but the realignment criterion is considerably affected. We could imagine that introducing additional noise cause faster decay of entanglement, nevertheless, in this case non-distillable entanglement showed much respect to additional noise compared with distillable entanglement.

Let us now consider the locally equivalent state $\sigma_\alpha$, which may be obtained from Eq.~(\ref{Eq:H-BE}) with local unitary operation $U = \mathbb{I}_3 \otimes \theta$, with $\theta = | 0 \rangle\langle 1 | + | 1 \rangle\langle 0 | + | 2 \rangle\langle 2 |$. Then we have 
\begin{eqnarray}
\sigma_\alpha = U \rho_\alpha U^\dagger = \frac{2}{7} \, |\tilde{\Psi}_+\rangle\langle \tilde{\Psi}_+| + \frac{\alpha}{7} \, \tilde{\sigma}_+ + \frac{5-\alpha}{7} \, \tilde{\sigma}_- \,,
\end{eqnarray}
where the separable states are $\tilde{\sigma}_+ = 1/3 (| 0,0 \rangle\langle 0,0 | + |1,2 \rangle\langle 1,2 | + | 2,1 \rangle\langle 2,1 |)$, and $\tilde{\sigma}_- = 1/3 (| 1,1 \rangle\langle 1,1 | + |2,0 \rangle\langle 2,0 | + | 0,2 \rangle\langle 0,2 |)$, and $|\tilde{\Psi}_+\rangle = 1/\sqrt{3} (|0,1\rangle + |1,0\rangle + |2,2\rangle)$ a maximally entangled state. The local unitary transformations do not effect the static entanglement, however they may have a profound influence on the future trajectory of entanglement \cite{Rau}. The possible negative eigenvalues of the time evolved partially transpose matrix are given as
\begin{eqnarray}
\eta_1 = \eta_2 &=& \frac{1}{42} \left(5-\sqrt{4 \, \alpha^2 - 20 \, \alpha + 25 + 16 \, \gamma ^2 \, \gamma_A^2 \, \gamma_B^2 } \right) , \nonumber \\
\eta_3 &=& \frac{1}{42} \left(5-\sqrt{4 \, \alpha^2 - 20 \alpha + 25 + 16 \, \gamma_A^4 \, \gamma_B^4 } \right) .
\end{eqnarray}
Again we observe that in the presence of only collective interaction, $\eta_3$ is always negative which implies no entanglement sudden death and no distillability sudden death. Whereas in the presence of either global decoherence or multi-local decoherence, the entanglement sudden death and/or distillability sudden death may happen.

\begin{figure}[h]
\scalebox{1.9}{\includegraphics[width=1.8in]{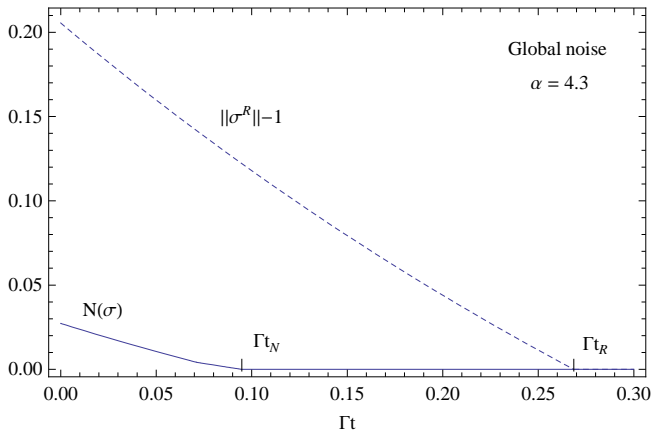}}
\caption{The negativity $N(\sigma)$ and the realignment criterion $\|\sigma^R\|-1$ is plotted against the decay parameter $\Gamma t$ for $\alpha = 4.3$. The state becomes PPT at an earlier time $\Gamma t_N \approx 0.0948$, nevertheless, it is entangled in the range $0.0948 \lesssim \Gamma t \lesssim 0.2686$.}
 \label{Fig:DSD-lehbe1}
 \end{figure}

In Figure \ref{Fig:DSD-lehbe1}, we have plotted the negativity and the realignment criterion against the decay parameter $\Gamma t$. We observe that distillability sudden death occurs in this case as well. The states becomes PPT at a time $\Gamma t_N \approx 0.0948$. Nevertheless, the states are bound-entangled in the interval $0.0948 \lesssim \Gamma t \lesssim 0.2686$. The state $\sigma_\alpha(t)$ is fragile compared with $\rho_\alpha(t)$ in the sense that $\sigma_\alpha(t)$ becomes bound-entangled at earlier times, i.\,e., distillability sudden death comes earlier as compared with dynamics of $\rho_\alpha(t)$ (compare with Figure \ref{Fig:DSD-hbe1}). However, in this case the realignment criterion detects bound-entangled states for a broader range.

\begin{figure}[h]
\scalebox{1.9}{\includegraphics[width=1.8in]{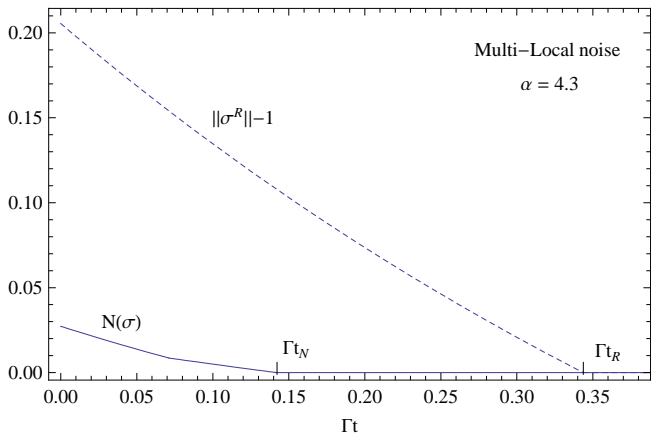}}
\caption{The negativity $N(\sigma)$ and the realignment criterion $\|\sigma^R\|-1$ is plotted against the decay parameter $\Gamma t$ for $\alpha = 4.3$. The state becomes bound-entangled at time $\Gamma t_N \approx 0.1422$, where negativity has zero value.}
\label{Fig:DSD-lehbe2}
\end{figure}

Figure \ref{Fig:DSD-lehbe2} shows the negativity $N(\sigma_\alpha(t))$ and $\|\sigma_\alpha^R(t)\|-1$ plotted against the decay parameter $\Gamma t$ for a specific value of parameter $\alpha$. We observe that $\sigma_\alpha(t)$ becomes bound-entangled at $\Gamma t \approx 0.1422$. Figure \ref{Fig:DSD-lehbe2} is exactly like Figure \ref{Fig:DSD-hbe2}, which means that the local unitary transformation has no effect on the dynamics of negativity and dynamics of bound-entangled states for this particular case.

Finally, let us consider the isotropic states defined as
\begin{eqnarray}
\rho_p = p \, |\Psi_+\rangle \langle \Psi_+| + \frac{(1-p)}{9} \mathbb{I}_9 \,, 
\end{eqnarray}
where $0 \leq p \leq 1$. The isotropic states have the property that their PPT region is always separable (see Ref. \cite{Clarisse} and references therein). Therefore, we realize that these states do not suffer distillability sudden death, however they may suffer entanglement sudden death. The possible negative eigenvalues of time evolved partially transpose matrix are given as
\begin{eqnarray}
\xi_1 &=& \frac{1}{9} (1 - p - 3 \, p \, \gamma^4 \, \gamma_A \, \gamma_B ) \,, \nonumber \\ 
\xi_2 &=& \frac{1}{9} \left(1 - p - 3 \, p \, \, \gamma^4 \, \gamma_A \, \gamma_B \right) \, , \nonumber \\ 
\xi_3 &=& \frac{1}{9} \left(1 - p - 3 \, p \, \gamma_A^2 \, \gamma_B^2 \right) \, .
\end{eqnarray}
We can see that isotropic states for $0 < p < 1$, do not suffer entanglement sudden death under collective decoherence. However in the presence of global decoherence or at least for multi-local noise, they always exhibit entanglement sudden death for $0 < p < 1$.

\section{Summary}\label{Sec:Conc}

We have revisited the entanglement dynamics of bipartite systems under global, collective and multi-local noises. In particular, we have studied the distillability sudden death of qutrit-qutrit systems. We have found that certain free-entangled qutrit-qutrit states becomes bound-entangled as a consequence of purely dynamical process. We have found that distillability sudden death may occur under the combined action of multi-local and collective noises. Distillability sudden death can also result under multi-local decoherence. We have observed that local unitary transformations can not avoid this phenomenon. Therefore, we conclude that qutrit-qutrit states evolve into three main types of dynamics: (i) They loose their entanglement only at infinity and hence never suffer entanglement sudden death and distillability sudden death. (ii) They loose their entanglement at a finite time but never undergo distillability sudden death. (iii) They first undergo distillability sudden death and then sudden death of entanglement.

\begin{acknowledgments}
I would like to thank Prof. Christof Wunderlich for his kind hospitality at Univesit\"at Siegen. I am also thankful to K. Chen for explaining his realignment criterion.
\end{acknowledgments}

\end{document}